# Measurement theory of a density profile of colloid particles on a flat surface: Conversion of force acting on a colloidal probe into pressure on its surface element

**Ken-ichi Amano**

*Department of Energy and Hydrocarbon Chemistry, Graduate School of Engineering, Kyoto University, Kyoto 615-8510, Japan.*

Author to whom correspondence should be addressed: Ken-ichi Amano.

Electric mail: amano.kenichi.8s@kyoto-u.ac.jp

**ABSTRACT**

Recently, we proposed a method that converts the force between two-large colloids into the pressure on the surface element (FPSE conversion) in a system of a colloidal solution. Using it, the density distribution of the small colloids around the large colloid is calculated. In a similar manner, in this letter, we propose a transform theory for colloidal probe atomic force microscopy (colloidal probe AFM), which transforms the force acting on the colloidal probe into the density distribution of the small colloids on a flat surface. If measured condition is proper one, in our view, it is possible for the transform theory to be applied for liquid AFM and obtain the liquid structure. The transform theory we derived is briefly explained in this letter.

**MAIN TEXT**

Recently, we proposed two methods for obtaining the density distribution of the small colloids around the large colloid [1,2]. In the method written in [2], the force acting on the large colloid is converted into the pressure on the surface element (FPSE conversion). Applying FPSE conversion, in this letter, a transform theory for colloidal probe AFM [3-6] is proposed. The purpose of the theory is determination of the density distribution of colloid particles on a flat surface. In our view, the transform theory is available for liquid AFM to determination of the liquid structure on a flat surface, if the condition is proper one.

From here, we derive the transform theory in the system of Fig. 1. There are the colloidal probe and the flat plate. The radius of the probe is $r_B$. The colloidal solution contains colloid particles with number density $\rho_0$. The radius of the colloid particle is $r_S$. The solution itself is inert background (ignored). A space which the center of the small colloid cannot enter is excluded volume of the colloidal probe, and $r$ is the radius of the excluded volume. (When $r_B \gg r_S$, $r \approx r_B$.) $\theta$ denotes the radian from the upward-vertical line originating from the center of the colloidal probe. The separation between the flat surface (right side) and the center of the colloidal probe is represented as $s$. The length of the horizontal line between the excluded surfaces of the flat plate and the colloidal probe is represented as $l$. If the force acting on the colloidal probe ($f$) is expressed by a summation of forces between face-to-face surface elements (see closed circles in Fig. 1), the force can be expressed as

$$f(s) = \sum_l P(l) A_{Pz}(l; s), \tag{1}$$

where $P$ is pressure and $A_{Pz}$ is an efficient area of the surface element of the colloidal probe, which is normal to $z$-axis. In the present case, Eq. (1) is rewritten as

$$f(s) = 2\pi r^2 \int_0^{\frac{\pi}{2}} P(l) \sin\theta \cos\theta d\theta. \tag{2}$$

It is considered that $P$ is "pressure from left" pluses "that from right sides" (see Fig. 1). In addition, we would like to mention that the force $f$ is colloid-induced mean force. That is, "$f$" plus "two-body force between the flat surface and the colloidal probe" is the directly measurable mean force (in terms of the inert solution). By the way, $l$ can be expressed as

$$l = s - r_S - r\sin\theta, \tag{3}$$

and hence following two expressions are obtained:

$$\cos\theta d\theta = -(1/r)dl, \tag{4}$$

$$\sin\theta = (s - r_S - l)/r. \tag{5}$$

Thus, Eq. (2) is rewritten as

$$f(s)/(2\pi) = \int_{s-r_S-r}^{s-r_S} P(l)(s - r_S - l)dl. \tag{6}$$

It can be seen that Eq. (6) is in the form of a matrix calculation as follows:

$$\boldsymbol{F}^* = \boldsymbol{HP}, \tag{7}$$

where $\boldsymbol{F}^*$ corresponds to left-hand side of Eq. (6). $\boldsymbol{P}$ and $\boldsymbol{H}$ correspond to $P(l)$ and the other parts, respectively. $\boldsymbol{H}$ is a square matrix whose variables are $l$ and $s$, however, its lower right area is composed of a square unit matrix. $\boldsymbol{P}$ is numerically calculated by using, for example, the inverse matrix of **H**. Consequently, $P(l)$ is obtained. The change from $f$ to $P$ of the surface element is FPSE conversion.

Density distribution of the colloid particles on the flat surface can be calculated by using the transform theories for SFA [7,8]. If the colloidal probe is approximated by a rigid one, the density distribution is obtained as follows [7]:

$$g_\text{F}(r_\text{S} + l) = \frac{P(l)}{k_\text{B}T\rho_0 g_\text{PC}} + 1, \tag{8}$$

where $g_\text{PC}$ and $g_\text{F}$ are the normalized number density of the colloid particles at the contact with the probe and the normalized number density of them on the flat surface, respectively. $k_\text{B}$ and $T$, respectively, are the Boltzmann constant and absolute temperature. The value of $g_\text{PC}$ is prepared by applying Asakura-Osawa theory (AO theory) [9-11] or measuring the force curve between chemically the same surfaces. Since the latter is explained in [2,7], the former is explained here. AO theory can estimate the change in free energy upon contact of the colloid particle at the colloidal probe from the change in translational space for the colloid particles ($\Delta V$). The volume change is calculated based on the excluded-volume change ($\Delta V > 0$). According to AO theory, the change in free energy being $\Delta F$ is expressed as

$$\Delta F \approx -T\Delta S = -k_B T \rho_0 \Delta V, \tag{9}$$

where $\Delta S$ is the change in entropy. Thus, $g_\text{PC}$ is given by [12]

$$g_\text{PC} = \exp[-\Delta F/(k_B T)] \approx \exp(\rho_0 \Delta V). \tag{10}$$

Even without the concrete value of $g_\text{PC}$, the *outline* of the density distribution can be estimated by Eq. (8), but it is better for us to obtain the more precise result. Although we introduced the complementary way for obtaining $g_\text{PC}$ above, AO theory is not so precise one. Therefore, the more precise estimation of $g_{PC}$ is our next study.

As a practical use of the transform theory, we transform a force curve measured in a nanoparticle suspension [13] into the number density distribution of the nanoparticles on a substrate. The nanoparticle-induced force is obtained by subtracting the bare force between the colloidal-probe and the substrate (Fig. 8(c) in [13]) from the net force ("N = 0" curve of Fig. 9 in [13]). Then, the nanoparticle-induced force is transformed into the number density distribution (see Fig. 2), where we set the volume fraction of the nanoparticle as 4.0 %, the diameter of the nanoparticle as 26 nm, the diameter of the colloidal-probe as 6.7 μm. In the transform, we approximated that the both surfaces of the colloidal-probe and the substrate are chemically the same and used a patch method [14], because the patch method outputs better results. Furthermore, we applied a smoothing function to the tail of the number density after 200 nm.

We have proposed the transform theory for calculating the density distribution of the colloid particles on the flat surface from the force curve obtained by colloidal probe AFM. The theory utilizes FPSE conversion [2] and the transform theory for surface force apparatus [7,8]. In our view, the transform theory (or basic concept of the transform theory) is useful for analyses of the density distributions of colloidal particles, nanoparticles, liquid molecules on substrates.

**ACKNOWLEDGEMENTS**

We appreciate Tetsuo Sakka (Kyoto University), Masahiro Kinoshita (Kyoto University), Hiroshi Onishi (Kobe University), Takeshi Fukuma (Kanazawa University), Naoya Nishi (Kyoto University), Kota Hashimoto (Kyoto University), Ryosuke Sawazumi (Kyoto University), and Taira Ishihara (Kyoto University) for the useful advice, discussions, and data. (Author K. Amano derived the transform theory and wrote its calculation program, K. Hashimoto and T. Ishihara extracted the numerical data from [13], and then we obtained Fig. 2.) This work was supported by5

"Grant-in-Aid for Young Scientists (B) from Japan Society for the Promotion of Science (15K21100)".

**FIGURES**

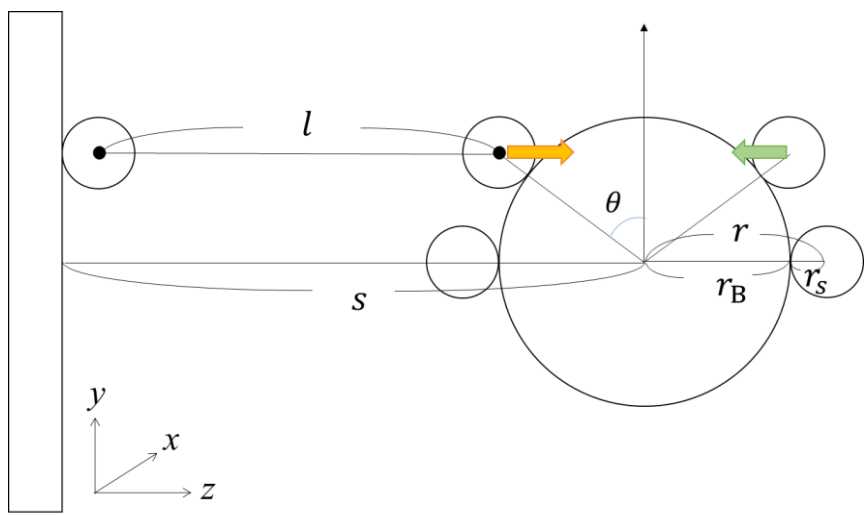

Fig. 1: The system configuration for the colloidal probe AFM. There are many small colloids with number density $\rho_0$. The solution is inert background.

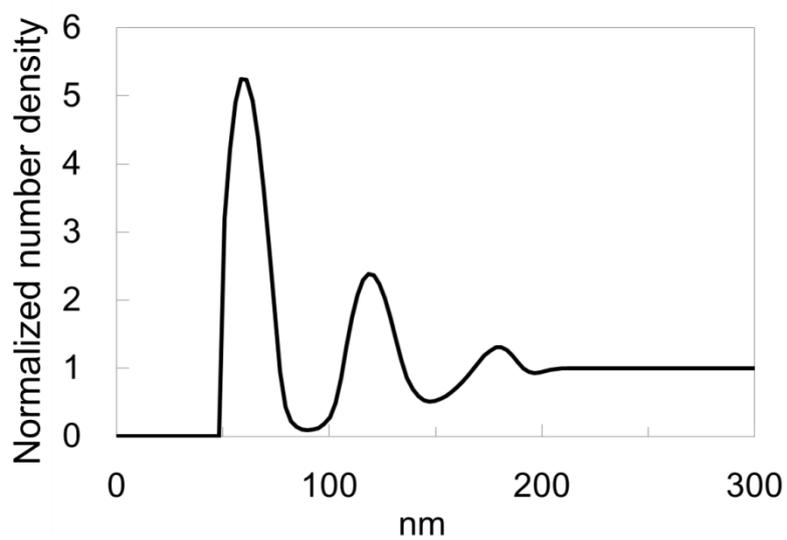

Fig. 2: Normalized number density distribution of the nanoparticles on the substrate.

7